\documentclass[a4paper,11pt,english]{article}
\usepackage{a4wide}
\usepackage{babel}
\usepackage{amssymb,amsmath}
\usepackage[small,bf,hang]{caption}
\usepackage{tikz}
\usepackage{tensor}
\usepackage[titles]{tocloft}
\usepackage{hyperref}
\usepackage[
	backref=true,
	sorting=none,
	maxnames=99,
	firstinits=true,
	doi=false
	]{biblatex}

\definecolor{darkred}{rgb}{0.7,0.15,0.15}
\definecolor{darkgreen}{rgb}{0.05,0.5,0.05}

\newcommand{\bg}{\bar{g}}
\newcommand{\bR}{\bar{R}}
\newcommand{\bcd}{\bar{\nabla}}
\newcommand{\cd}{{\nabla}}

\newcommand{\pertn}[1]{#1^L}

\newcommand{\za}{\mu}
\newcommand{\zb}{\nu}
\newcommand{\zc}{\rho}
\newcommand{\zd}{\sigma}

\DeclareMathOperator{\sgn}{sgn}

\def\ddal{\mathop{\vrule height 7pt depth0.2pt
\hbox{\vrule height 0.5pt depth0.2pt width 6.2pt}\vrule height 7pt depth0.2pt width0.8pt
\kern-7.4pt\hbox{\vrule height 7pt depth-6.7pt width 7.pt}}}
\def\sdal{\mathop{\kern0.1pt\vrule height 4.9pt depth0.15pt
\hbox{\vrule height 0.3pt depth0.15pt width 4.6pt}\vrule height 4.9pt depth0.15pt width0.7pt
\kern-5.7pt\hbox{\vrule height 4.9pt depth-4.7pt width 5.3pt}}}
\def\ssdal{\mathop{\kern0.1pt\vrule height 3.8pt depth0.1pt width0.2pt
\hbox{\vrule height 0.3pt depth0.1pt width 3.6pt}\vrule height 3.8pt depth0.1pt width0.5pt
\kern-4.4pt\hbox{\vrule height 4pt depth-3.9pt width 4.2pt}}}

\def\dal{\mspace{1.5mu}\mathchoice{\ddal}{\ddal}{\sdal}{\ssdal}\mspace{1.5mu}}

\begin{document}

\pagestyle{empty}
\hfill AEI-2012-030 \phantom{l.....}
\vskip 0.12\textheight
{\bfseries\Huge{\noindent Polycritical Gravities}}
\vfill

{\bfseries\Large{\noindent Teake Nutma}}
\\%
\\%
\phantom{.}\begin{tabular}{cp{0.8\textwidth}}
& \emph{Max-Planck-Institut f\"ur Gravitationsphysik}\\
& \emph{(Albert Einstein Institut)}\\
& \emph{Am M\"uhlenberg 1, 14476 Golm, Germany}\\ 
& \\
& \emph{Email:}
\href{mailto:teake.nutma@aei.mpg.de}{\nolinkurl{teake.nutma@aei.mpg.de}}
\end{tabular}
\\ 
\\
\\
\\
{\bfseries\Large{\noindent Abstract}}\\
\\
\phantom{.}\begin{tabular}{cp{0.9\textwidth}}
& We present higher-derivative gravities that propagate an
arbitrary number of gravitons of different mass on (A)dS backgrounds. These
theories have multiple critical points, at which the masses degenerate and the
graviton energies are non-negative. For six derivatives and higher there are 
critical points with positive energy.
\end{tabular}
\\
\\
\\
\\
\begin{minipage}{0.945\textwidth}
\tableofcontents
\end{minipage}

\newpage
\pagestyle{plain}

\section{Introduction}

Two-derivative Einstein gravity in four dimensions is non-renormalizable. It
can be made perturbatively renormalizable by adding four-derivative terms
to the Lagrangian \cite{Stelle:1976gc,Stelle:1977ry}. However, the
addition of the curvature-squared terms spoils unitarity: around a Minkowski
background they introduce a massive spin-0 and spin-2 mode. These massive modes
have norm opposite of the massless spin-2 mode, and thus are ghosts. The spin-0
mode can be eliminated by tuning the coefficients of the curvature-squared
terms, but the massive spin-2 modes cannot.

Recently, a consistent four-derivative theory of gravity in three
dimensions, called `New Massive Gravity' was introduced in
\cite{Bergshoeff:2009hq}. NMG is ghost-free due to the fact that massless
gravitons have no propagating degrees of freedom in three dimensions,
which makes it possible to choose the overall sign of the action such that the
massive gravitons have positive energy. This, however, is not possible in higher dimensions as there both massive and massless gravitons propagate.

One way around this problem is to perturb around an (A)dS background,
instead of a Minkowski background. The cosmological constant and the
coefficients of the curvature squared terms can then be tuned such that the
massive modes becomes massless \cite{Liu:2009bk}. This is known as `critical
gravity' \cite{Lu:2011zk}. As the massive modes disappear at the critical
point, the theory is potentially unitary. 

However, at the critical point the massive modes are replaced by so-called log
modes \cite{Grumiller:2008qz,Liu:2009kc,Bergshoeff:2011ri}. As it turns out,
these log modes are ghosts \cite{Porrati:2011ku}, and must the truncated to restore unitarity.
As their falloff in the radial AdS coordinate is logarithmic (hence the name),
this may be done by imposing certain boundary conditions.

The resulting theory is then unitary, but, unfortunately, also empty. Namely,
at the critical point the energy of the massless graviton modes vanishes,
together with the mass of the Schwarzschild black hole.
It was recently argued from a CFT perspective \cite{Bergshoeff:2012sc} that this
is essentially due to the fact that critical gravity is of rank two (with the
rank being half the number of maximum derivatives). Instead, gravity theories of
odd rank should not suffer from this `zero-energy-problem'.

The purpose of this paper is to investigate the criticality conditions for
higher-rank theories of gravity. It is organized as follows. We first give a
non-linear Lagrangian for arbitrary rank $r$, that, on (A)dS backgrounds,
propagates one massless and $r-1$ massive gravitons, but not the scalar ghost
mode.
Next, we show that the quadratic perturbation of this Lagrangian and its linear
equations of motion can be concisely written in terms of the so-called
\emph{Schouten operator}. This reformulation enables us to calculate the global
charges (such as black hole masses) and graviton energies for arbitrary
rank. From the latter we deduce that the theory is critical, i.e.~all
energies are non-negative, whenever sufficiently enough graviton masses are
degenerate. In general there will be more than one critical point; hence the
name \emph{polycritical} gravities.

\section{Non-linear action}
\label{sec:startingpoint}

For a gravity theory of rank $r$ (thus containing at most $2r$
derivatives), we would like its linear equations of motion to be\footnote{See
\autoref{app:conventions} for our conventions on linearization.}
\begin{subequations}
\label{eq:eom1expanded}
\begin{align}
	\prod_{n=0}^{r-1} \left( \bar{\dal} - 2\Lambda - m^2_n \right)
	h_{\za\zb} & = 0 , 				\label{eq:tteom1} 		\\[-10pt]
	\bcd^\za h_{\za\zb} & = 0 , 		\label{eq:ttgauge1} 	\\
	\bg^{\za\zb} h_{\za\zb} & = 0 . 	\label{eq:ttgauge2}
\end{align}
\end{subequations}
This a straightforward  generalization of the Fierz-Pauli equations of motion
for a single massive graviton \cite{Hinterbichler:2011tt}. Here however we have
$r$ graviton modes with a priori different masses $m_0$, $m_1, \ldots, m_{r-1}$.
Because these equations of motion should follow from some covariant non-linear
Lagrangian, one of the masses (say $m_0$) will always be zero. This is due to
the diffeomorphism invariance of the non-linear theory. We will use the
index $n = 0, 1, \ldots , r-1$ to indicate all gravitons, and the index
$i = 1, \ldots, r-1$ for only the massive gravitons.

In four dimensions and higher, a non-linear Lagrangian
whose linearized equations of motion are those given above, is
\begin{equation}
\label{eq:action_ansatz}
	\mathcal{L} = \sqrt{-g} \left[
		R -  (d-2)(d-1) \Lambda
		+ \frac{1}{4} C_{\za\zb\zc\zd} \left(\sum_{i=1}^{r-1} a_i \dal^{i-1} \right)
		C^{\za\zb\zc\zd} \right] .
\end{equation}
Here $C$ denotes the Weyl tensor, and the coefficients $a_i = a_i
(r,d,\Lambda,m_i)$ are functions of the rank, the dimension, the cosmological
constant, and the graviton masses. For $r=2$, this action was already written
down in \cite{Deser:2011xc}. We will give explicit values for the coefficients
$a_i$ for $r = 3$ below. 

Note that we use the canonical sign for the Einstein-Hilbert
term in the above action. Flipping its sign is equivalent to changing the overall
sign of the action, upon redefining $\Lambda$ and $a_i$ accordingly. Such a change
of sign also changes the sign of the energy of the solutions (see 
\autoref{sec:graviton_energies}), and, as noted in the introduction, is particularly
important in the $d=3$, $r=2$ case \cite{Bergshoeff:2009hq}. There it is customary 
to leave the sign of the Einstein-Hilbert term arbitrary. Here, however, we have 
fixed the sign, keeping in mind that we can always flip the overall sign of the
action in order to choose which modes have positive energy and which negative.

We have two main reasons for using only Weyl tensors in the higher order
terms. Both stem from the fact that the Weyl tensor vanishes identically on
(A)dS spaces. First, this ensures the uniqueness of the (A)dS vacuum.
Second, for perturbations around such a background the higher order terms do
not contribute to the trace of the equations of motion. This comes about as
follows.

The full non-linear equations of motion that follow from
\eqref{eq:action_ansatz} are
\begin{equation}
\label{eq:eom}
 	E_{\za\zb} 
 		= -\frac{1}{\sqrt{-g}}\frac{\delta \mathcal{L}}{\delta g^{\za\zb}} 
 		= G_{\za\zb} + \sum_{i=1}^{r-1} a_i K^i_{\za\zb} 
 		= 0 .
\end{equation}
Here $G_{\za\zb}$ is the
cosmological Einstein tensor (see \autoref{app:conventions}), and $K^i_{\za\zb}$
are the contributions from the higher-order terms. Suppressing indices on the Weyl
tensor, these contributions have the generic form
\begin{equation}
\label{eq:deltaCboxC}
	K^i_{\za\zb} = 
		 2 \frac{\delta C}{\delta g^{\za\zb}}  \dal^i C
		+ C \left( \frac{\delta \dal^i}{\delta g^{\za\zb}}  - \frac{1}{2}g_{\za\zb}
		\right) C .
\end{equation}
Thus $K^i_{\za\zb}$ consists of a part that is linear in the Weyl tensor, and
part that is quadratic. For an (A)dS space, both parts are zero. So on these
backgrounds just the Einstein-Hilbert contribution of \eqref{eq:eom} survives.
This uniquely fixes the background curvature to be $\Lambda$. 

For linear perturbations around (A)dS solutions, the part of $K^i_{\za\zb}$ that
is quadratic in the Weyl tensor vanishes. The linearized higher-order
contributions to $E^L_{\za\zb}$ come then only from the first term on the right-hand side of \eqref{eq:deltaCboxC}, which
evaluates to
\begin{equation}
	\left(K^i_{\za\zb}\right)^L 
	=  \bcd^\zc \bcd^\zd {\bar\dal}^i C^L_{\za\zc\zb\zd} .
\end{equation}
The linear Weyl tensor $C^L$ is, just like its non-linear variant, traceless.
Upon taking the trace of the linear equations of motion, it follows that the
linear Ricci scalar vanishes on-shell:
\begin{equation}
\label{eq:eom1trace}
	\bg^{\za\zb} E^L_{\za\zb} = \left( 1 - \frac{d}{2} \right) \pertn{R} = 0 .
\end{equation}
As in Einstein gravity, this allows us to impose the transverse traceless gauge
\cite{Wald:1984rg}, i.e.~equations \eqref{eq:ttgauge1} and \eqref{eq:ttgauge2},
for the linear graviton fluctuations $h_{\za\zb}$. Hence the scalar mode $h$, which would
otherwise be a ghost, does not propagate.

In the remainder of this section we will show that the linear equations of
motion take the form \eqref{eq:tteom1}, and give explicit values of the Lagrange
parameters $a_i$ for the rank $r=3$. 
The linearized equations of motion can be written entirely in terms of
$G^L_{\za\zb}$ and $R^L$ by using the identities
\begin{subequations}
\label{eq:w1identities}
\begin{align}
	\bcd^\zc \bcd^\zd \bar{\dal}^i  C^L_{\za\zc\zb\zd} & = 
		\Bigl[ \bar{\dal} + 2(d-2)\Lambda \Bigr]^i \, \bcd^\zc \bcd^\zd
		C^L_{\za\zc\zb\zd} ,\\
	 \bcd^\zc \bcd^\zd C^L_{\za\zc\zb\zd} & =
	 \frac{d-3}{d-2} \Bigl[ \bar{\dal} - d \Lambda  \Bigr]  G^L_{\za\zb} 
	 + 
	   \frac{1}{2} \frac{d-3}{d-1} \Bigl[
	 	\bg_{\za\zb} \left( \bar{\dal}- (d-1)\Lambda \right)- \bcd_\za \bcd_\zb 
	 \Bigr]  R^L .
\end{align}
\end{subequations}
The former follows from commuting covariant derivatives, while the latter is a
consequence of the Bianchi identities. Furthermore, in the transverse
traceless gauge the linear Einstein tensor takes on the form
\begin{equation}
\label{eq:einsteintt}
	G^L_{\za\zb} 
	= -\frac 1 2 \left( \bar{\dal} - 2 \Lambda \right) h_{\za\zb} .
\end{equation}
And as $R^L = 0$ on-shell, the linear equation of motion $E^L_{\za\zb}$ is a
polynomial in $\bar\dal$ that acts on $h_{\za\zb}$. We may always choose the
parameters $a_i$ such that it factorizes into the form \eqref{eq:tteom1}.
Indeed, for $r=3$ the linear equation of motion becomes
\begin{equation}
	E^L_{\za\zb} = -\frac{1}{2\tau}
		\bigl( \bar\dal - 2 \Lambda \bigr)
		\prod_{i=1}^{r-1} \left( \bar\dal - 2 \Lambda - m^2_i \right)
	h_{\za\zb} = 0 ,
\end{equation}
where the parameter $\tau$ is given by
\begin{equation}
\label{eq:tau}
	\tau = \prod_{i=1}^{r-1} \Bigl[ m^2_i - (d-2)\Lambda \Bigr] ,
\end{equation}
and the squared masses by
\begin{subequations}
\begin{align}
	m^2_1 & = - \frac{d \Lambda}{2} - \frac{a_1 + \sqrt{b} }{2 a_2} , \\ 
	m^2_2 & = - \frac{d \Lambda}{2} - \frac{a_1 - \sqrt{b} }{2 a_2}  , \\
	b	& = \Bigl[ a_1 + (3d - 4) \Lambda a_2 \Bigr]^2 - 4 \frac{d-2}{d-3} a_2 .
\end{align}
\end{subequations}
Inverting the above equations for $a_0$ and $a_1$ gives finally
\begin{subequations}
\begin{align}
	a_1 & = -\frac{1}{\tau} \frac{d-2}{d-3} (m^2_1 + m^2_2 + d \Lambda) , \\ 
	a_2 & = +\frac{1}{\tau} \frac{d-2}{d-3} .
\end{align}
\end{subequations}
The factors of $d-3$ in the denominator indicate that the non-linear Lagrangian
\eqref{eq:action_ansatz} is only valid for $d \geq 4$. This makes sense, as the
Weyl tensor vanishes identically in three dimensions and lower. Similar explicit
values for the parameters $a_i$ of higher-rank theories can be computed along
the same lines. While these explicit values are needed for the non-linear
action, they are not for its quadratic perturbation. As we will see in
the next section, the latter can be written concisely in
closed form using the mass parameters $m_i$ instead of the parameters $a_i$.
Furthermore, the quadratic Lagrangian will also be valid in three dimensions.

\section{Quadratic action}

Before we set out to calculate the conserved charges and energies of our
higher-rank theory, it is convenient to rewrite linear equations of motion a
bit. We start by rearranging the quadratic perturbation of
the non-linear Lagrangian \eqref{eq:action_ansatz}. It is given by
\begin{equation}
	\mathcal{L}_2 = 
		- \frac{1}{2} h^{\za\zb} G^L_{\za\zb} 
		- \frac{1}{2} \bcd^\zd \bcd^\zb h^{\za\zc} \left( \sum_{i=1}^{r-1} a_i\,
		{\bar\dal}^{i-1} \right) C^L_{\za\zb\zc\zd} .   
\end{equation}
We have dropped a total derivative in the Einstein part, and expanded one of the
linear Weyl tensors in terms of the graviton fluctuations $h_{\za\zb}$. There
are more contributions to this expansion than $\bcd^\zd \bcd^\zb h^{\za\zc}$,
but they drop out because of the contraction with the other Weyl tensor.

Like the linear equations of motion, the quadratic Lagrangian can be written
entirely in terms of $G^L_{\za\zb}$ and $R^L$ by using the identities
\eqref{eq:w1identities}. The resulting expression can be simplified further to
\begin{equation}
\label{eq:l2schouten}
	\mathcal{L}_2 = -\frac{1}{2\tau} G_L^{\za\zb} \left( 
		\prod_{i=1}^{r-1}  2 \mathcal{S} + m^2_i
		\right) \circ h_{\za\zb} .
\end{equation}
Here $\tau$ is given as in \eqref{eq:tau}, and we have introduced the
\emph{Schouten operator} $\mathcal{S}$.
It is defined such that when it acts on the graviton
fluctuations $h_{\za\zb}$, it gives the linear cosmological Schouten tensor:
\begin{equation}
\label{eq:schoutendef}
	\mathcal{S} \circ h_{\za\zb} 
	\equiv S^L_{\za\zb} .
\end{equation}
The cosmological Schouten tensor $S_{\za\zb}$ is in turn defined such that for
vanishing $\Lambda$ it reduces to the normal Schouten tensor, and that
it is zero on (A)dS backgrounds, i.e.~$\bar{S}_{\za\zb} = 0$. See also
\autoref{app:conventions}.
Surprisingly, the quadratic action \eqref{eq:l2schouten} is also valid in three
dimensions, whereas the non-linear action \eqref{eq:action_ansatz} was not. For
$d=3$, $r=2$ and $\Lambda = 0$ it coincides with the
quadratic action given in \cite{Bergshoeff:2009fj}.

Before deriving equations of motion from \eqref{eq:l2schouten}, we first list
some useful properties of the Schouten operator and the Einstein operator
$\mathcal G$.
The latter is defined in a similar fashion as the Schouten operator,
\begin{equation}
\label{eq:einsteindef}
	\mathcal{G} \circ h_{\za\zb} 
	\equiv G^L_{\za\zb} .
\end{equation}
For arbitrary symmetric tensors $A_{\za\zb}$ and $B_{\za\zb}$, we have
\begin{subequations}
\begin{align}
	B^{\za\zb} \, \mathcal{S} \circ A_{\za\zb} & = 
		A^{\za\zb} \biggl[
		\mathcal{S} \circ B_{\za\zb} 
		+ \frac{1}{2}\frac{d-2}{d-1} \Bigl( 
			\bcd_\za \bcd_\zb B
			- \bg_{\za\zb} \bcd^\zc \bcd^\zd B_{\zc\zd} 
		\Bigr)
		\biggr] , \label{eq:SGident_Spart} \\
	B^{\za\zb} \, \mathcal{G} \circ A_{\za\zb} & = 
		A^{\za\zb} \, \mathcal{G} \circ B_{\za\zb} , \\[8pt]
	\bcd^\za \, \mathcal{G} \circ A_{\za\zb} & = 0 , \\
	\bigl[ \mathcal{G}, \mathcal{S} \bigr]  A_{\za\zb} & = 
		\frac{1}{2} \frac{d-2}{d-1} \bcd_\za \bcd_\zb \, \bg^{\zc\zd} \, \mathcal{G}
		\circ A_{\zc\zd} . \label{eq:SGident_comm}
\end{align}
\end{subequations}
In the first two lines we have dropped a total derivative while integrating by
parts. Note that for $B_{\za\zb} = \mathcal G \circ A_{\za\zb}$ the last term
of \eqref{eq:SGident_Spart} vanishes, and the middle term is the same as 
the commutator \eqref{eq:SGident_comm}. Thanks to this subtle interplay between
the Schouten and Einstein operators, the linear equation of motion reads
\begin{equation}
\label{eq:eom1schouten}
	E^L_{\za\zb} 
		= - \frac{\delta \mathcal{L}_2}{\delta h^{\za\zb}}
		= \frac{1}{\tau} \, \mathcal{G} \circ  \left( 
		\prod_{i=1}^{r-1}  2 \mathcal{S} + m^2_i
		\right) \circ h_{\za\zb} .
\end{equation}  
Upon taking the trace of these equations, we should recover
\eqref{eq:eom1trace}, that is, $R^L = 0$. To see how this comes about, we need
three additional properties of the Schouten and Einstein operators:
\begin{subequations}
\label{eq:SGidents2}
\begin{align}
	\bg^{\za\zb} \, \mathcal{G} \circ A_{\za\zb} & =
		-(d-1) \bg^{\za\zb} \, \mathcal{S} \circ A_{\za\zb} , \\
	\bg^{\za\zb} \, \mathcal{S} \circ A_{\za\zb} & =
		\frac{1}{2}\frac{d-2}{d-1} \Bigl[
			\bcd^\za \bcd^\zb A_{\za\zb} -  \bar\dal A  - (d-1) \Lambda A
		 \Bigr] , \\
	\bcd^\za \, \mathcal{S} \circ A_{\za\zb} & =
		\bcd_\zb \, \bg^{\zc\zd} \, \mathcal{S} \circ A_{\zc\zd} ,
\end{align}
\end{subequations}
from which it follows that
\begin{equation}
	\bg^{\za\zb} \, \mathcal{G} \circ \mathcal{S} \circ  A_{\za\zb} =
	- \frac{d-2}{2}\Lambda \,\bg^{\za\zb} \, \mathcal{G}  \circ  A_{\za\zb} .
\end{equation}
A short calculation shows that we indeed recover \eqref{eq:eom1trace}:
\begin{equation}
\label{eq:eom1schoutentrace}
	\bg^{\za\zb} E^L_{\za\zb} 
		 = \frac{1}{\tau} \, \bg^{\za\zb}   \left( 
		\prod_{i=1}^{r-1} m^2_i - (d-2)\Lambda
		\right) \,\mathcal{G} \circ h_{\za\zb} 
		 = \left(1 - \frac{d}{2}\right) R^L = 0 . 
\end{equation}  
As the linear Ricci scalar vanishes on-shell, we may go to the
transverse-traceless gauge \eqref{eq:ttgauge1}, \eqref{eq:ttgauge2}. In this
gauge the Schouten and Einstein operators become equal (compare equation
\eqref{eq:einsteintt}),
\begin{equation}
\label{eq:SGtt}
	\mathcal S \circ h_{\za\zb} = \mathcal G \circ h_{\za\zb} 
	= -\frac 1 2 \left(	\bar{\dal} - 2 \Lambda \right) h_{\za\zb} .
\end{equation}
The complete linear equations of motion \eqref{eq:eom1schouten} can then be
written as
\begin{subequations}
\label{eq:eom1final}
\begin{align}
	\frac{(-1)^r}{2\tau} \prod_{n=0}^{r-1} \bigl( \bar{\dal} - 2\Lambda - m^2_n
	\bigr) h_{\za\zb} & = 0 , 				\label{eq:eom1allmodes}		\\[-10pt]
	\bcd^\za h_{\za\zb} & = 0 , 	 	\\
	\bg^{\za\zb} h_{\za\zb} & = 0 ,	
\end{align}
\end{subequations}
with $m_0=0$.

\section{Conformal invariance}

The overall factor $\tfrac{1}{\tau}$ in the quadratic Lagrangian
\eqref{eq:l2schouten} comes
from demanding that the Ricci scalar in the non-linear action
\eqref{eq:action_ansatz} has the usual normalization. The advantage of this
normalization is that we recover Einstein gravity upon decoupling the
massive gravitons by sending their the masses to infinity. We will see later
in sections \ref{sec:charges} and \ref{sec:graviton_energies} that also the
conserved charges and graviton energies reduce to their two-derivative
`Einstein' values in this limit.

However, an obvious drawback of the overall factor $\tfrac{1}{\tau}$ is that it
has poles at the mass values
\begin{equation}
\label{eq:conformal_point}
	m^2_i = (d-2)\Lambda .
\end{equation}
%
One easy way to get rid of the poles is to simply replace the overall factor
$\tfrac{1}{\tau}$ by some other factor $\tfrac{1}{\tau '}$ that has the same
mass dimension, but no explicit dependence on $m_i$ and thus no poles. We can
then freely let the masses take the values \eqref{eq:conformal_point}, with the
drawback that we do not recover Einstein gravity upon decoupling the massive
gravitons. Another possible drawback could be that for the mass values 
\eqref{eq:conformal_point} the trace of the linear equations of motion
\eqref{eq:eom1schoutentrace} vanishes identically, and does not eliminate the
scalar mode of the graviton.

Luckily, the latter does not happen. Instead, for the values
\eqref{eq:conformal_point} the linear theory develops a conformal invariance. To
see how this happens, consider the linear conformal
transformation
\begin{equation}
	\delta_\omega h_{\za\zb} = \bg_{\za\zb} \, \omega .
\end{equation}
We would like to know the variation of the equations of motion
\eqref{eq:eom1schouten} under this transformation. To this end we first compute
the variation of a single Schouten operator,
\begin{equation}
\label{eq:conf_schouten}
	\delta_\omega \left(\mathcal S \circ h_{\za\zb}\right)   
	= \mathcal S \circ \left( \bg_{\za\zb} \, \omega \right) 
	= - \frac{d-2}{2}
	\left( \Lambda \bg_{\za\zb} +\bcd_\za \bcd_\zb \right) \omega .
\end{equation}
Next, we notice the identities
\begin{equation}
	\mathcal G  \circ \left(  \bcd_\za \bcd_\zb \, \omega \right)  
	= \mathcal S \circ \left( \bcd_\za \bcd_\zb \, \omega \right) = 0.
\end{equation}
Thus for the repeated composition of the Schouten operator only the first
term on the right-hand side of \eqref{eq:conf_schouten} is important. The
variation of the equations of motion \eqref{eq:eom1schouten} then becomes
\begin{equation}
	\delta_\omega E^L_{\za\zb} 
		 = \frac{1}{\tau} \left( 
		\prod_{i=1}^{r-1} m^2_i - (d-2)\Lambda
		\right) \,\delta_\omega \left( \mathcal{G} \circ h_{\za\zb} \right) .
\end{equation} 
This is zero for the mass values \eqref{eq:conformal_point} and the redefinition
of $\tau$ mentioned above. This makes it possible to choose the conformal gauge
$\omega = \tfrac{h}{d}$, such that the scalar mode vanishes everywhere.

The extra conformal gauge symmetry is somewhat reminiscent of the `partially
massless' modes that occur in two-derivative Fierz-Pauli theory
\cite{Deser:2001pe,Deser:2001us}. At the critical value of the
Higuchi bound Fierz-Pauli theory also develops an extra gauge symmetry
\cite{Deser:1983mm}, although not a conformal one. The extra gauge symmetry for
the higher-derivative theories considered here can be thought of as a
generalization of the two-derivative partially massless case.

\section{Conserved charges}
\label{sec:charges}

We now derive the conserved charges of our theory. They can be
calculated via the Abbott-Deser method \cite{Abbott:1981ff}, which is an
extension of the ADM energy \cite{Arnowitt:1962hi,Regge:1974zd} to backgrounds
with constant curvature. In this method the linearized equations of motion
$E^L_{\za\zb}$ are treated as an effective energy-momentum tensor. This allows
us to compute conserved charges $Q^\za$ as follows:
\begin{equation}
\label{eq:ad_charges}
	Q^\za (\bar \xi) = \int_{\Sigma} \textrm{d}^{d-1} x \, \sqrt{-\bg} \,
	E_L^{\za\zb} \, \bar\xi_\zb .
\end{equation}
Here $\bar\xi_\za$ is a Killing vector of the background, and $\Sigma$ is a
spatial $(d-1)$ dimensional hypersurface. For instance, the global mass of a
solution is then given by $Q_0$ for a time-like Killing vector. The trick for calculating
the conserved charges is to show that the integrand can be written as a
divergence of a two-form $F_{\za\zb}$,
\begin{equation}
	E_L^{\za\zb} \, \bar\xi_\zb = \bcd_\zb F^{\za\zb} .
\end{equation}
The integral in \eqref{eq:ad_charges} then reduces to a surface integral at
spatial infinity,
\begin{equation}
\label{eq:surfaceintegral}
	Q^\za (\bar \xi) = \int_{\partial\Sigma} dS_\alpha \, F^{\za\alpha} ,
\end{equation}
where $\partial\Sigma$ is the $(d-2)$ dimensional boundary of $\Sigma$.
For Einstein-Hilbert gravity, whose linear equation of motion is simply
$\mathcal G \circ h_{\za\zb} = 0$, the two-form is
\begin{align} 
	F^{\textrm{EH}}_{\za\zb} 
	& = 	\bar\xi^\zc \bcd_{[\za} h_{\zb]\zc}
		+ \bar\xi_{[\za} \bcd_{\zb]} h
		- \bar\xi_{[\za} \bcd^\zc h_{\zb]\zc}
		+ h^\zc{}_{[\za} \bcd_{\zb]} \bar\xi_\zc
		+ \frac{1}{2} h \bcd_{\za}\bar\xi_\zb  \nonumber \\
	& \equiv \mathcal{F}_{\bar\xi} \circ h_{\za\zb} .
\end{align}
Here we have introduced the two-form operator $\mathcal{F}_{\bar\xi}$. From the
definition above, we have the following property. When it acts on a
symmetric tensor, it gives a two-form whose divergence is the contraction of
the Einstein operator with a Killing vector:
\begin{equation}
	\bcd^\zb \, \mathcal{F}_{\bar\xi} \circ A_{\za\zb} = \bar\xi^\zb \, \mathcal G
	\circ A_{\za\zb} .
\end{equation} 
In our case the linear equation of motion is \eqref{eq:eom1schouten}.
Its general structure is the same as that of Einstein gravity, namely a symmetric
tensor hit by the Einstein operator. Hence the corresponding two-form
simply reads
\begin{equation}
\label{eq:twoform_schouten}
	F_{\za\zb} 
		= \frac{1}{\tau} \, \mathcal{F}_{\bar\xi} \circ  \left( 
		\prod_{i=1}^{r-1} 2 \mathcal{S} + m^2_i
		\right) \circ h_{\za\zb} .
\end{equation}
Like \cite{Deser:2002rt,Deser:2002jk} we restrict to solutions that are
asymptotically (A)dS. That is, at spatial infinity the vacuum
Einstein equations are satisfied:
\begin{equation}
	G^L_{\za\zb} \Big|_{\partial\Sigma} = 0,\qquad  
	R^L  \Big|_{\partial\Sigma} = 0 , \qquad
	S^L_{\za\zb} \Big|_{\partial\Sigma} = 0. 
\end{equation}
The last equation follows from the fact that the linear Schouten tensor can be
decomposed as $S^L_{\za\zb} = G^L_{\za\zb} + \tfrac{1}{2}\tfrac{d-2}{d-1}
\bg_{\za\zb} R^L$. So the terms with Schouten operators in
\eqref{eq:twoform_schouten} are zero in the asymptotic region, and all that
remains is the product of the squared masses $m^2_i$. Suppressing the dependency on the
Killing vector, we obtain
\begin{equation}
\label{eq:charge}
	Q^\za 
	= \frac{Q_{\textrm{EH}}^\za}{\tau}  \prod_{i=1}^{r-1} m^2_i .
\end{equation}
Thus the conserved charges are equal to those of two-derivative
Einstein-Hilbert gravity, up to a renormalization factor. In the limit when all
extra graviton modes become infinitely heavy and decouple, the renormalization
factor goes to one by \eqref{eq:tau}.  Furthermore the conserved charges vanish
when one of the graviton masses is zero, which is what happens at the critical point in
four-derivative critical gravity \cite{Lu:2011zk,Deser:2011xc}.

\section{Graviton energies}
\label{sec:graviton_energies}

In this section we will derive the energies associated to the different graviton
modes $h^{(n)}_{\za\zb}$. These modes are annihilated by a single
factor of the product in the complete equation of motion
\eqref{eq:eom1allmodes},
\begin{equation}
\label{eq:graviton_modes}
	\left( \bar\dal - 2 \Lambda - m^2_n \right) h^{(n)}_{\za\zb} = 0 .
\end{equation}
In \cite{Li:2008dq,Liu:2009bk,Lu:2011zk} the graviton energies were computed by
first deriving the Hamiltonian from an effective action. The Hamiltonian was then
evaluated on-shell for the different graviton modes in order to give the energy.
The disadvantage of this approach is that one has to make an ADM-like split of
the indices and variables, and use the Ostrogradsky method to deal with the
higher derivatives.

Here we will follow a different route, as outlined in \cite{Iyer:1994ys},
that circumvents these inconveniences.
First we compute the energy-momentum tensor by varying the quadratic action
\eqref{eq:l2schouten} with respect to the background metric:
\begin{equation}
	T_{\za\zb} = 
		2 \frac{\delta \mathcal{L}_2}{\delta \bg^{\za\zb}} .
\end{equation}
The energy is then obtained by integrating this energy-momentum tensor
$T_{\za\zb}$ over a Cauchy surface,
\begin{equation}
	\mathcal E = \int_\Sigma \textrm{d}^{d-1}x \, T_{\za\zb} n^\za \bar\xi^\zb .
\end{equation}
Here $n^\za$ is the unit normal to $\Sigma$ and  $\bar\xi^\zb$ is a time-like
Killing vector.

For Einstein-Hilbert gravity, the on-shell energy-momentum tensor is
\begin{equation}
	T^\textrm{EH}_{\za\zb} = 
		 - h^{\zc\zd} \frac{\delta G^L_{\zc\zd}}{\delta \bg^{\za\zb}} 
		 = - h^{\zc\zd} \frac{\delta \mathcal G}{\delta \bg^{\za\zb}} \circ h_{\zc\zd}
		 .
\end{equation}
For deducing the energy-momentum tensor of our theory we need one last
identity involving the Schouten and Einstein operators. First note from
\eqref{eq:SGidents2} that when the Schouten and Einstein operators act on a
transverse and traceless tensor, the resulting tensor is also transverse and
traceless. Furthermore, by equation \eqref{eq:SGtt}, their action on transverse
traceless tensors gives the same result. This implies that for arbitrary
transverse traceless symmetric tensors $A_{\za\zb}$ and $B_{\za\zb}$, we have
\begin{align}
	B^{\za\zb} \delta_{\bg} \left(  \mathcal S \circ
	A_{\za\zb} \right) 
		& = B^{\za\zb}  (\delta_{\bg} \mathcal G) \circ A_{\za\zb}
		+ \left( \delta_{\bg} A_{\za\zb} \right) \, \mathcal{G} \circ B^{\za\zb} .
\end{align}
Because we will evaluate the energy-momentum tensor on-shell, $h_{\za\zb}$ is
transverse and traceless by \eqref{eq:eom1final}. Thus we are allowed to use the
above identities in deriving the energy-momentum tensor. Lastly, from equations
\eqref{eq:SGtt} with \eqref{eq:graviton_modes}, we have on-shell 
\begin{equation}
	\mathcal S \circ h^{(n)}_{\za\zb} = \mathcal G \circ h^{(n)}_{\za\zb}
	= -\frac{1}{2} m^2_n h^{(n)}_{\za\zb} .
\end{equation} 
Combining the above equations, the energy-momentum tensor becomes
\begin{equation}
	T^{(n)}_{\za\zb} = 
		\frac{1}{\tau} \Biggl[ 
			\prod_{i=1}^{r-1} (m^2_i - m^2_n)
			- m^2_n \sum_{i=1}^{r-1} \prod_{\substack{j=1\\j\neq i}}^{r-1} (m^2_j - m^2_n) 
		\Biggr]
		T^{\textrm{EH}}_{\za\zb} .
\end{equation}
The superscript $(n)$ indicates that it is evaluated on-shell for the mode
$h^{(n)}_{\za\zb}$. When $m_n = 0$, only the first product contributes to the
energy-momentum tensor. For non-zero masses $m_i$, the first product vanishes and only
one term in the sum is non-zero. This gives the following energies for the
massless and massive gravitons:
\begin{subequations}
\label{eq:graviton_energies}
\begin{align}
	\mathcal{E}^{(0)} & = +\frac{\mathcal{E}_\textrm{EH}}{\tau} 
		\prod_{j=1}^{r-1} m^2_j 
	   \\
	\mathcal{E}^{(i)} & = - \frac{\mathcal{E}_\textrm{EH}}{\tau} m^2_i 
		\prod_{\substack{j=1\\j\neq i}}^{r-1}( m^2_j - m^2_i ) 
	  .
\end{align}
\end{subequations}
The energy of the massless graviton $h^{(0)}_{\za\zb}$ has the same overall
factor as the conserved charge $Q^\za$ \eqref{eq:charge}. So it seems that the
massive gravitons $h^{(i)}_{\za\zb}$ do not contribute to the conserved charge.
This is to be expected, as the massive gravitons fall of too fast towards
spatial infinity to contribute to the surface integral
\eqref{eq:surfaceintegral} in the asymptotic region.

For four-derivative theories ($r=2$), the above energies reduce to
$\mathcal{E}^{(0)} = - \mathcal{E}^{(1)}$, which matches with the energies
found in \cite{Liu:2009bk, Lu:2011zk}. So for rank two the only way to obtain
energies with the same sign is to set the mass $m_1$ to zero. Both energies are then zero, and the
conserved charge also vanishes, rendering the theory trivial.

In the six-derivative case, $r=3$, this zero-energy problem does not occur. The
graviton energies \eqref{eq:graviton_energies} then namely read
\begin{subequations}
\begin{align}
	\mathcal{E}^{(0)} & = \frac{\mathcal{E}_\textrm{EH}}{\tau}m^2_1 m^2_2    ,\\
	\mathcal{E}^{(1)} & = \frac{\mathcal{E}_\textrm{EH}}{\tau}m^2_1 (m^2_1 - m^2_2)  ,\\
	\mathcal{E}^{(2)} & = \frac{\mathcal{E}_\textrm{EH}}{\tau}m^2_2 (m^2_2 - m^2_1)  .
\end{align}
\end{subequations}
\begin{figure}[t]
\centering
\begin{tikzpicture}[domain=-3*pi/8:7*pi/8,scale=2]
   \draw[thin,->] (-1.55,0) -- (3.3,0) node[right] {$\theta$};
   \draw[thin,->] (-1.5,-0.7) -- (-1.5,1.7) node[above]
   {$\mathcal{E}^{(i)}$ $\left[\frac{\mathcal{E}_\textrm{EH}}{\tau}\right]$};
   \draw[color=darkred]  plot[smooth] (\x,{sin(\x r)*sin(\x r) - sin(\x r)*cos(\x r)})
       node[right] {$\mathcal{E}^{(2)}$};
   \draw[color=blue] plot[smooth] (\x,{sin(\x r)*cos(\x r)})
       node[right] {$\mathcal{E}^{(0)}$};
   \draw[color=darkgreen] plot[smooth] (\x,{cos(\x r)*cos(\x r) - sin(\x
   r)*cos(\x r)}) node[right] {$\mathcal{E}^{(1)}$};
   \draw[fill=white,color=white] (-0.7,-0.485) circle(0.2) {};
   \draw[fill=white,color=white] (2.465,-0.475) circle(0.2) {};
    \small{
   \draw (-1.55,-1/2) node[anchor=east] {$-1$};
   \draw (-1.55,1/2) node[anchor=east] {$1$};
   \draw (-1.55,1) node[anchor=east] {$2$};
   \draw (-1.55,0) node[anchor=east] {$0$};
   \draw (0,-0.1) node[anchor=north west,rotate=-45] {$m^2_2=0$};
   \draw (pi/4,-0.1) node[anchor=north west,rotate=-45] {$m^2_1 = m^2_2$}; ;
   \draw (pi/2,-0.1) node[anchor=north west,rotate=-45] {$m^2_1=0$};
   \draw (-pi/4,-0.1) node[anchor=north west,rotate=-45] {$m^2_1=-m^2_2$};
   \draw (3*pi/4,-0.1) node[anchor=north west,rotate=-45] {$m^2_1=-m^2_2$};
   };
   \draw[thin]  (-1.55,-1/2) -- (-1.5,-1/2);
   \draw[thin]  (-1.55,1/2) -- (-1.5,1/2);
   \draw[thin]  (-1.55,1) -- (-1.5,1);
   \draw[thin]  (0,0) -- (0,-0.07);
   \draw[thin]  (pi/4,0) -- (pi/4,-0.07);
   \draw[thin]  (-pi/4,0) -- (-pi/4,-0.07);
   \draw[thin]  (pi/2,0) -- (pi/2,-0.07);
   \draw[thin]  (3*pi/4,0) -- (3*pi/4,-0.07);
\end{tikzpicture}
\caption{Graviton energies for $r=3$ in polar coordinates. Here 
$\theta=\tan^{-1} \left(\frac{m^2_2}{m^2_1}\right)$ is
the angle in the $(m^2_1,m^2_2)$ plane. The
masses are given by $m^2_1 = 2\cos \theta$ and $m^2_2 = 2\sin \theta$. There
are three points where all energies are non-negative: $m^2_1 =
m^2_2$, $m^2_1 = 0$, or $m^2_2 = 0$.}
\label{fig:energies}
\end{figure}
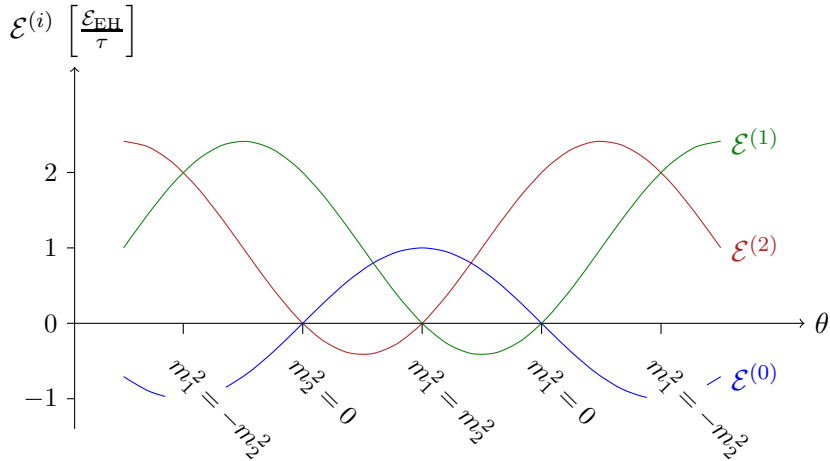%
These energies are plotted in \autoref{fig:energies}. 
There are two distinct points where they have same sign: either when $m^2_1 = m^2_2$
or when $m^2_1 = 0 \vee m^2_2 = 0$. In the last critical point the conserved is
zero, whereas it can be positive in the first.

For yet higher rank theories there are even more critical points.
However, it will never be possible to have the same sign for all energies without
degeneracies in the masses. If we namely arrange the masses by size,
\begin{equation}
\label{eq:mass_arrange}
	m^2_1 < \ldots < m^2_i < \ldots < m^2_{r-1} ,
\end{equation}
the sign of the energies \eqref{eq:graviton_energies} alternates:
\begin{equation}
	\sgn\left(\mathcal{E}^{(i)}\right) = - \sgn\left(\mathcal{E}^{(i+1)}\right) .
\end{equation}
Thus the situation with all masses different \eqref{eq:mass_arrange}
leads to ghosts; to avoid this we need at least some degeneracy of
the masses.

But whenever there is a mass $m_n$ with multiplicity $\mu_n > 1$, a so-called
log-mode $h_{\za\zb}^{(n , p)}$ appears \cite{Bergshoeff:2011ri}.
These log-modes are annihilated not by a single, but by multiple factors of the
product in the equations of motion,
\begin{equation}
	\bigl( \bar\dal - 2 \Lambda - m^2_n \bigr)^p h^{(n,p)}_{\za\zb} = 0 .
\end{equation}
The label $p$ can take the values $p = 2, \ldots, \mu_n$, as $p=1$ simply gives
the non-logarithmic graviton mode $h_{\za\zb}^{(n)}$. 
From the four-derivative case the log-modes are expected to be ghosts
\cite{Porrati:2011ku}, and, if possible, need to be truncated out in order to
restore unitarity. 

\section{Conclusions}

In this paper we have studied gravities of arbitrary rank, meaning they
propagate any number of gravitons on (A)dS backgrounds. Besides from giving a
quadratic and a non-linear action, we have calculated the conserved charges and
the graviton energies. From the energies we deduce that there will 
be ghosts unless the masses have critical values. At these critical points some
of the gravitons have degenerate mass.
But as mass degeneracies lead to logarithmic graviton modes, the untruncated
theory will never be unitary. By truncating the log-modes by imposing
appropriate boundary conditions one could obtain a unitary sub-sector of the theory. We leave the
exact form of both the higher-rank log-modes and boundary conditions to future
study.

When the rank is two, there is only one critical point, and all the energies
vanish \cite{Liu:2009bk, Lu:2011zk,Deser:2011xc}. One can interpret the
triviality of this theory as being to due to the proposed equivalence of Einstein gravity and
conformal gravity \cite{Maldacena:2011mk,Lu:2011ks}. We have shown that for
higher rank theories there are critical points where the conserved charges and
graviton energies do not vanish. But in the fully degenerate case where all the
graviton masses are zero, the theory will always be empty. Like the proposed
equivalence of Einstein and conformal gravity, this `emptiness' of higher rank
theories could in principle be used to construct a chain of equivalence
relations between gravity theories of different rank.

\appendix

\section{Conventions}
\label{app:conventions}

We use the `mostly plus' metric signature $(-,+,\ldots,+)$. The conventions for
the Riemann tensor are the default of the \emph{xAct} software package
\cite{xAct:2012}, which in turn follows Wald's conventions \cite{Wald:1984rg}:
\begin{align}
	[\cd_\za,\cd_\zb] T_\zc & = R_{\za\zb\zc}{}^\zd T_\zd , &
	R_{\za\zb} & = R_{\za\zc\zb}{}^\zc .
\end{align}
Barred objects are background quantities (i.e.~$\bg$ denotes the background
metric). AdS and dS backgrounds are chosen as follows:
\begin{subequations}
\label{eq:adsbackground}
\begin{align}
\bR_{\za\zb\zc\zd}  & = 
	\Lambda \left( \bg_{\za\zc} \bg_{\zb\zd} -
		\bg_{\za\zd} \bg_{\zb\zc} \right) ,  \\
\bR_{\za\zb}	& = 
	(d-1) \Lambda \bg_{\za\zb} , \\
\bR & = 
	d (d-1)  \Lambda .
\end{align}
\end{subequations}
Perturbations around these backgrounds are defined as
\begin{equation}
	g_{\za\zb} = \bar g_{\za\zb} + g^L_{\za\zb} = \bar g_{\za\zb} + h_{\za\zb} .
\end{equation}
The superscript $L$ indicates linear perturbations. Thus the linear perturbation
of the metric is given by $h_{\za\zb}$.

The cosmological Einstein tensor is the usual Einstein tensor plus a term
proportional to the cosmological constant, such that it vanishes on the above
backgrounds:
\begin{subequations}
\begin{align}
	G^\Lambda_{\za\zb} & = R_{\za\zb} -\frac 1 2 g_{\za\zb} \Bigl[ R -
	(d-2)(d-1)\Lambda \Bigr] ,\\
	\bar{G}^\Lambda_{\za\zb} & = 0 .
\end{align}
\end{subequations}
The cosmological Schouten tensor is defined similarly,
\begin{subequations}
\begin{align}
	S^\Lambda_{\za\zb} & = R_{\za\zb} - \frac{1}{2} g_{\za\zb} \left[
		 \frac{R}{d-1} + (d-2) \Lambda	
	\right]  ,\\
	\bar{S}^\Lambda_{\za\zb} & = 0 .
\end{align}
\end{subequations}
The Schouten tensor is usually given with an additional overall factor
$\tfrac{1}{d-2}$. However, for our purposes the above definition is more
convenient. In the main text the superscripts $\Lambda$ are dropped from the
cosmological Einstein and Schouten tensors. Thus by $G_{\za\zb}$ and
$S_{\za\zb}$ we always mean their cosmological versions.

For completeness, we give the linear perturbations of the cosmological Einstein
and Schouten tensors:
\begin{subequations}
\begin{align}
	G^L_{\za\zb} & = R^L_{\za\zb}  - (d-1)\Lambda
	h_{\za\zb} - \frac{1}{2} \bg_{\za\zb} R^L ,	\\
	S^L_{\za\zb} & = R^L_{\za\zb}  - (d-1)\Lambda
	h_{\za\zb}- \frac{1}{2(d-1)} \bg_{\za\zb} R^L ,
\end{align}
\end{subequations}
with
\begin{subequations}
\begin{align}
	R^L_{\za\zb} & = \bcd_\zc \bcd_{(\za} h_{\zb)}{}^\zc  - \frac{1}{2} \bar\dal
	h_{\za\zb}  - \frac{1}{2} \bcd_\za \bcd_\zb h  , \\
	R^L			& =  \bcd_\zc \bcd_\zd
	h^{\zc\zd} - \bar\dal h - (d-1) \Lambda h .
\end{align}
\end{subequations}

\section*{Acknowledgements}

We would like to thank Sjoerd de Haan, Axel Kleinschmidt, Ilarion
Melnikov, Wout Merbis, and Stefan Theisen for useful
discussions. As some of the calculations of this paper have been carried out
using \emph{xAct} \cite{xAct:2012}, we thank Jose Martin-Garcia for making this
rather excellent piece of software freely available. Lastly, we thank Amitabh
Virmani and Andrea Campoleoni for discussions and feedback on early drafts.

\printbibliography

\end{document}